\documentclass[useAMS,usenatbib]{mn2e}
\usepackage{graphicx,natbib}
\usepackage{psfrag}
\usepackage{amssymb}
\usepackage{float}
\usepackage[dvips]{color}
\PassOptionsToPackage{linktocpage}{hyperref}

\title[Galactic spiral structure]{Tracing the Galactic spiral structure with embedded clusters}

\author[D. Camargo, C. Bonatto  and E. Bica]{D. Camargo$^{1,2}$, C. Bonatto$^1$ and E. Bica$^1$\\
$^1$ Departamento de Astronomia, Universidade Federal do Rio Grande do Sul, 
Av. Bento Gon\c{c}alves 9500\\
Porto Alegre 91501-970, RS, Brazil\\
$^2$ Col{\'e}gio Militar de Porto Alegre, Minist{\'e}rio da Defesa - Ex{\'e}rcito Brasileiro, 
Av. Jos{\'e} Bonif{\'a}cio 363\\
Porto Alegre 90040-130, RS, Brazil}

\begin{document}

\pagerange{\pageref{firstpage}--\pageref{lastpage}}

\maketitle

\label{firstpage}

\begin{abstract}
In the present work we investigate the properties of 18 embedded clusters (ECs). The sample includes 11 previously known clusters and we report the discovery of 7 ECs on WISE images, thus complementing our recent list of 437 new clusters.
The main goal is to use such clusters to shed new light on the Galactic structure by tracing the spiral arms with cluster distances. Our results favour a four-armed spiral pattern tracing three arms, Sagitarius-Carina, Perseus, and the Outer arm.
The Sagitarius-Carina spiral arm is probed in the borderline of the third and fourth quadrants at a distance from the Galactic centre of  $d_1\sim6.4$ kpc adopting $R_{\odot}=7.2$ kpc, or $d_2\sim7.2$ kpc for $R_{\odot}=8.0$ kpc.
Most ECs in our sample are located in the Perseus arm that is traced in the second and third quadrants and appear to be at Galactocentric distances in the range $d_1=9-10.5$ kpc or  $d_2=9.8-11.3$ kpc.
Dolidze 25, Bochum 2, and Camargo 445 are located in the Outer arm that extends along the second and third Galactic quadrants with a distance from the Galactic centre in the range of $d_1=12.5-14.5$ kpc or $d_2=13.5-15.5$ kpc.
We find further evidence that in the Galaxy ECs are predominantly located within the thin disc and along spiral arms. They are excellent tools for tracing these Galactic features and therefore new searches for ECs can contribute to a better understanding of the Galactic structure.
We also report an EC aggregate located in the Perseus arm.
\end{abstract}


\section{Introduction}
\label{Intro}

Despite the effort that has been made to improve our understanding of the Galactic structure, questions about the spiral arm nature \citep{Baba09, Sellwood11, Martinez13, Sellwood14, Grand12, Roca13}, structure \citep{Georgelin76, Russeil03, Levine06, Majaess09, Hou14}, and dynamics \citep{Fujii11, Binney14} remain open. There is no consensus on the number, pitch angle, and shape of Galactic spiral arms \citep{Vallee05, Hou09, Lepine11, Siebert12, Francis12, Vallee14a, Vallee14b, Griv14, Bobylev14, Pettitt14}. The Sun's location within the dust obscured Galactic disc is a complicating factor to observe the Galactic structure.

It is widely accepted that spiral arms are the preferred sites of star formation and, as most stars form within embedded cluster (EC) the arms are sites of cluster formation. Star formation may occur after the collapse and fragmentation of giant molecular clouds (GMCs) that occur within spiral arms transforming dense gas clumps into ECs. Based on the absence of massive $^{13}CO$ bright molecular clouds in the interarm space, \citet{Duval09} argue that molecular clouds must form in spiral arms and be short-lived (less than 10Myr). Then, the spiral arms may be traced by young star clusters, especially ECs that have not had enough time to move far from their birth places. In addition, EC parameters are derived with good accuracy. In this sense, ECs with derived parameters can be used to distinguish between the various theoretical models for spiral arm structure \citep{Dobbs10, Sellwood10}. 
Besides constraining  the spiral arm distribution  with direct distances, ECs can also contribute to kinematic modelling of the spiral structure \citep{Georgelin76, Russeil03}.

\begin{figure*}
\begin{minipage}[b]{1.0\linewidth}
\includegraphics[scale=0.36,angle=0,viewport=0 300 2000 1300,clip]{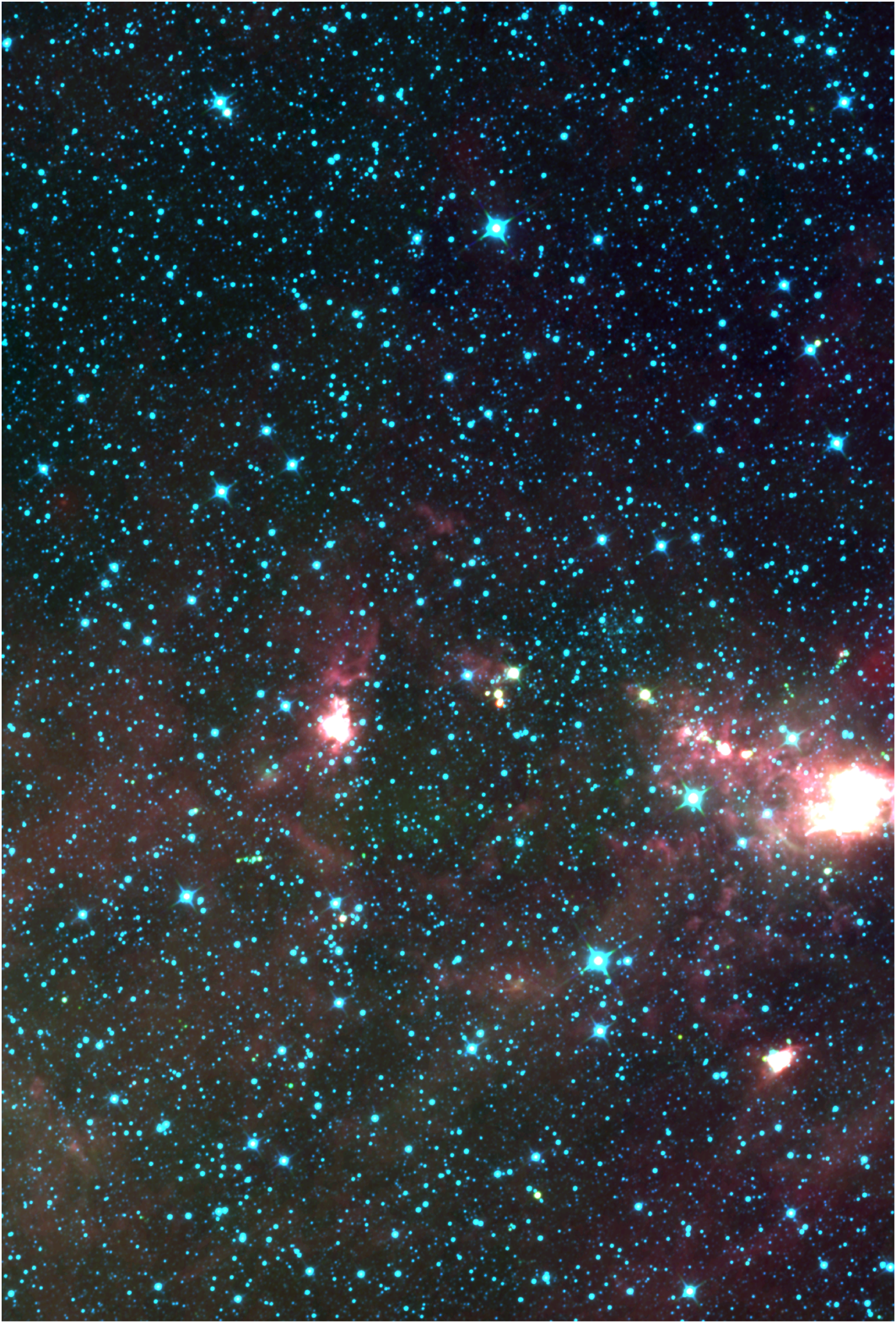}
\put(-380.0,300.0){\makebox(0.0,0.0)[5]{\fontsize{12}{12}\selectfont \color{yellow}FSR 665}}
\put(-530.0,138.0){\makebox(0.0,0.0)[5]{\fontsize{12}{12}\selectfont \color{yellow}C 443}}
\put(-585.0,165.0){\makebox(0.0,0.0)[5]{\fontsize{12}{12}\selectfont \color{yellow}C 444}}
\put(-440.0,270.0){\makebox(0.0,0.0)[5]{\fontsize{12}{12}\selectfont \color{yellow}FSR 666}}
\put(-250.0,275.0){\makebox(0.0,0.0)[5]{\fontsize{12}{12}\selectfont \color{yellow}C 441}}
\put(-250.0,210.0){\makebox(0.0,0.0)[5]{\fontsize{12}{2}\selectfont \color{yellow}BDS 61}}
\put(-320.0,235.0){\makebox(0.0,0.0)[5]{\fontsize{12}{12}\selectfont \color{yellow}C 442}}
\put(-530.0,255.0){\makebox(0.0,0.0)[5]{\fontsize{12}{12}\selectfont \color{yellow}BDS 63}}
\put(-290.0,60.0){\makebox(0.0,0.0)[5]{\fontsize{12}{12}\selectfont \color{yellow}BDS 62}}
\end{minipage}\hfill
\caption[]{WISE RGB image of the Perseus 1 aggregate: FSR 665, FSR 666, Camargo 441, Camargo 442,  Camargo 443, Camargo 444, BDS 61, BDS 62, and BDS 63.}
\label{nuvem}
\end{figure*}

We have contributed significantly to increase the number of ECs with derived parameters \citep{Camargo09, Camargo10, Camargo11, Camargo12, Camargo13, Camargo15a}. 
Besides deriving parameters, in \citet{Camargo15a} we discovered $437$ ECs and stellar groups increasing the Galactic EC sample. \citet{Majaess13} has also made a significant contribution on young clusters \citep[see also,][]{Bica03a, Bica03b, Lada03, Mercer05, Froebrich07, Borissova11, Solin14}.

\begin{figure*}

\begin{minipage}[b]{0.328\linewidth}
\includegraphics[width=\textwidth]{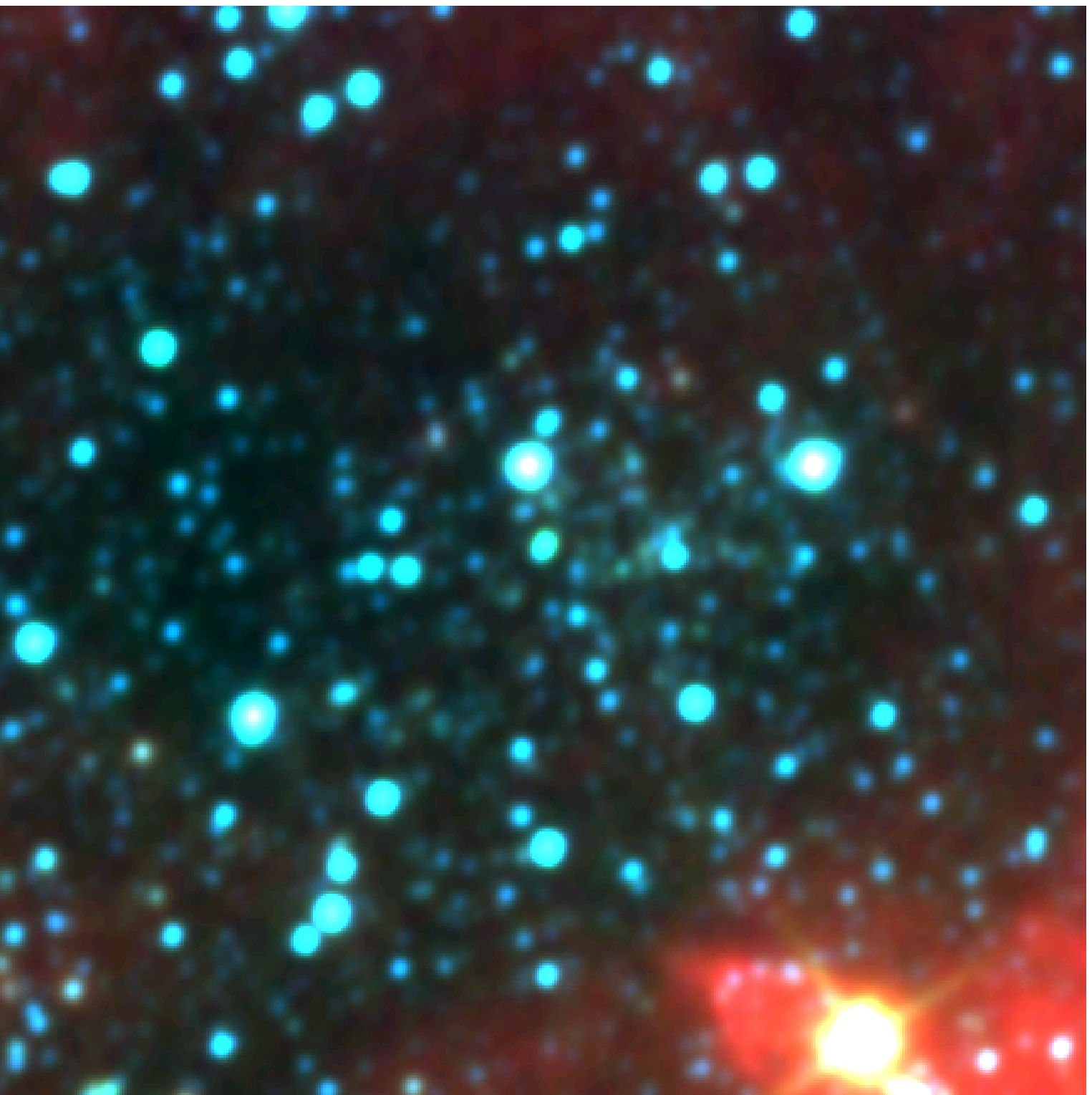}
\put(-134.0,155.0){\makebox(0.0,0.0)[5]{\fontsize{14}{14}\selectfont \color{red}FSR 665}}
\end{minipage}\hfill
%\hspace{0.01cm}
\vspace{0.02cm}
\begin{minipage}[b]{0.328\linewidth}
\includegraphics[width=\textwidth]{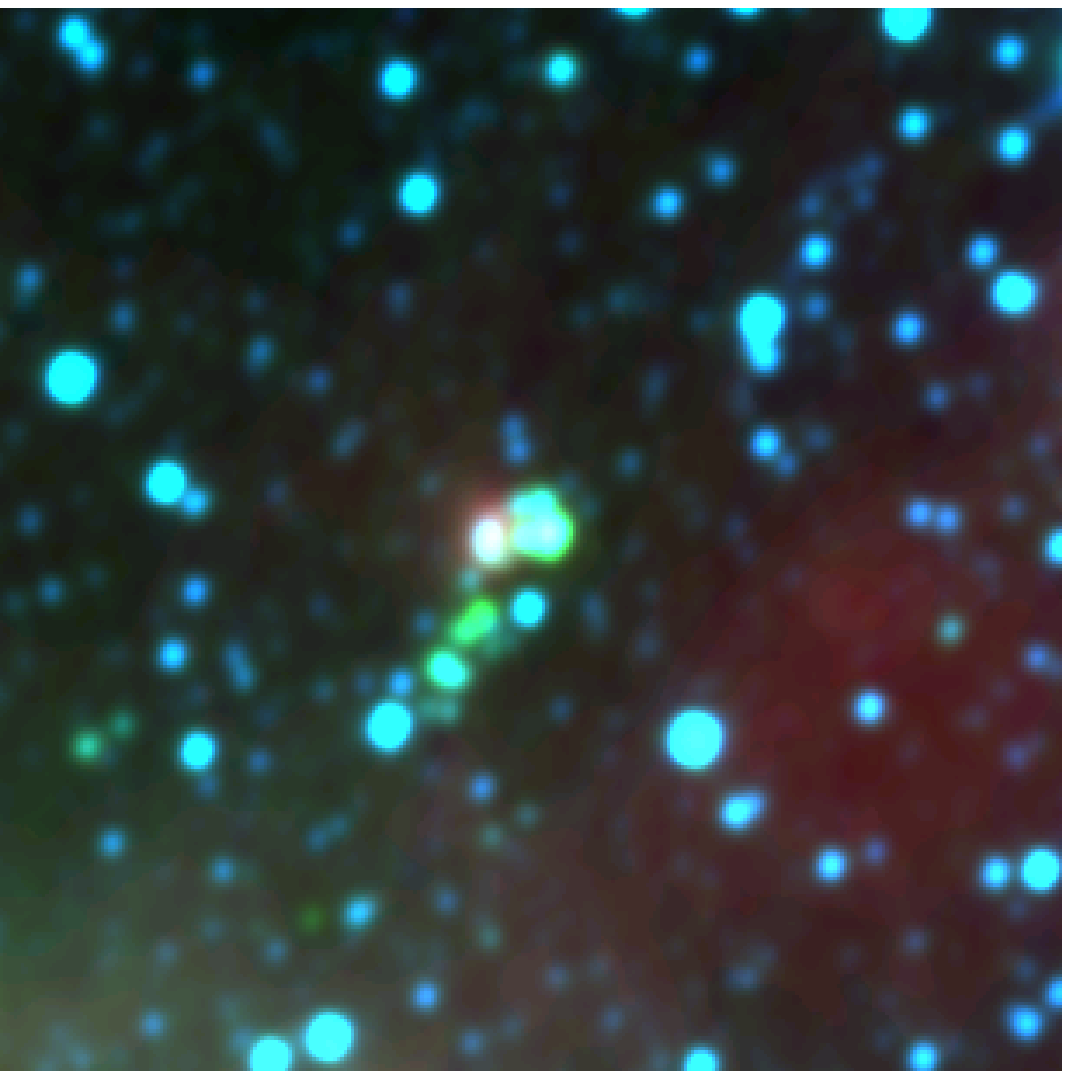}
\put(-140.0,155.0){\makebox(0.0,0.0)[5]{\fontsize{14}{14}\selectfont \color{red}C 441}}
\end{minipage}\hfill
\vspace{0.02cm}
\begin{minipage}[b]{0.328\linewidth}
\includegraphics[width=\textwidth]{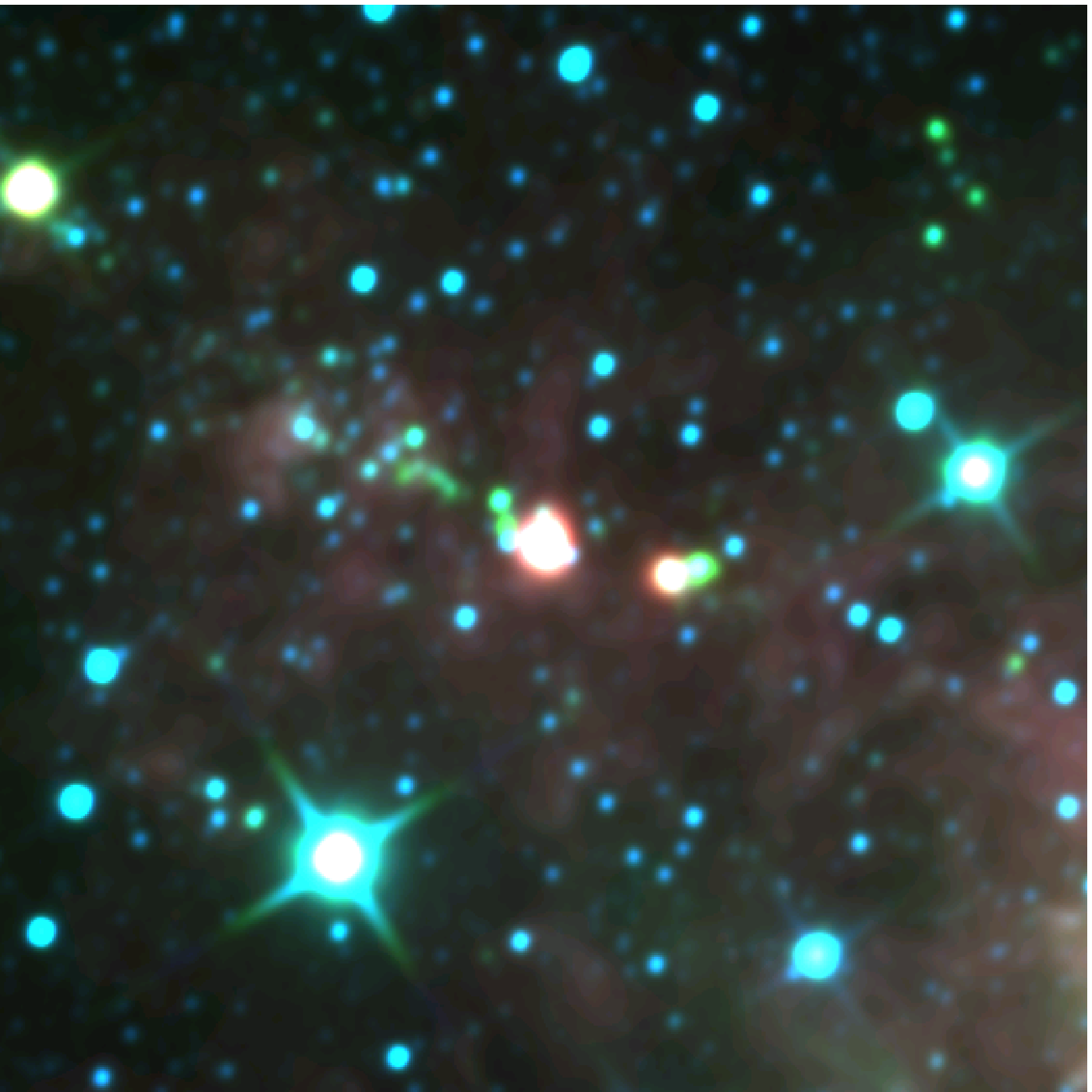}
\put(-140.0,155.0){\makebox(0.0,0.0)[5]{\fontsize{14}{14}\selectfont \color{red}C 442}}
\end{minipage}\hfill
%\hspace{0.01cm}
\vspace{0.02cm}
\begin{minipage}[b]{0.328\linewidth}
\includegraphics[width=\textwidth]{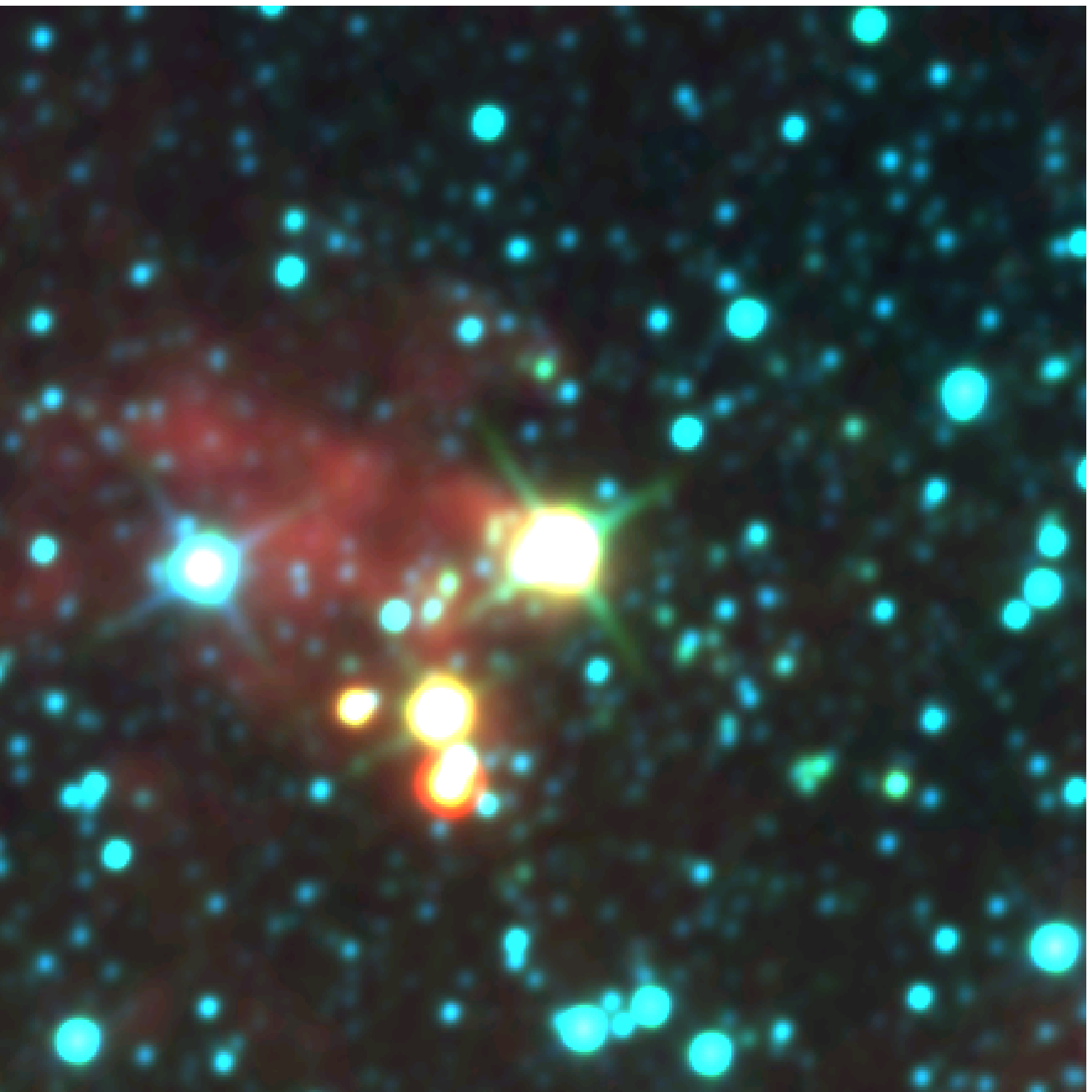}
\put(-137.0,155.0){\makebox(0.0,0.0)[5]{\fontsize{14}{14}\selectfont \color{red}FSR 666}}
\end{minipage}\hfill
%\hspace{0.01cm}
\begin{minipage}[b]{0.328\linewidth}
\includegraphics[width=\textwidth]{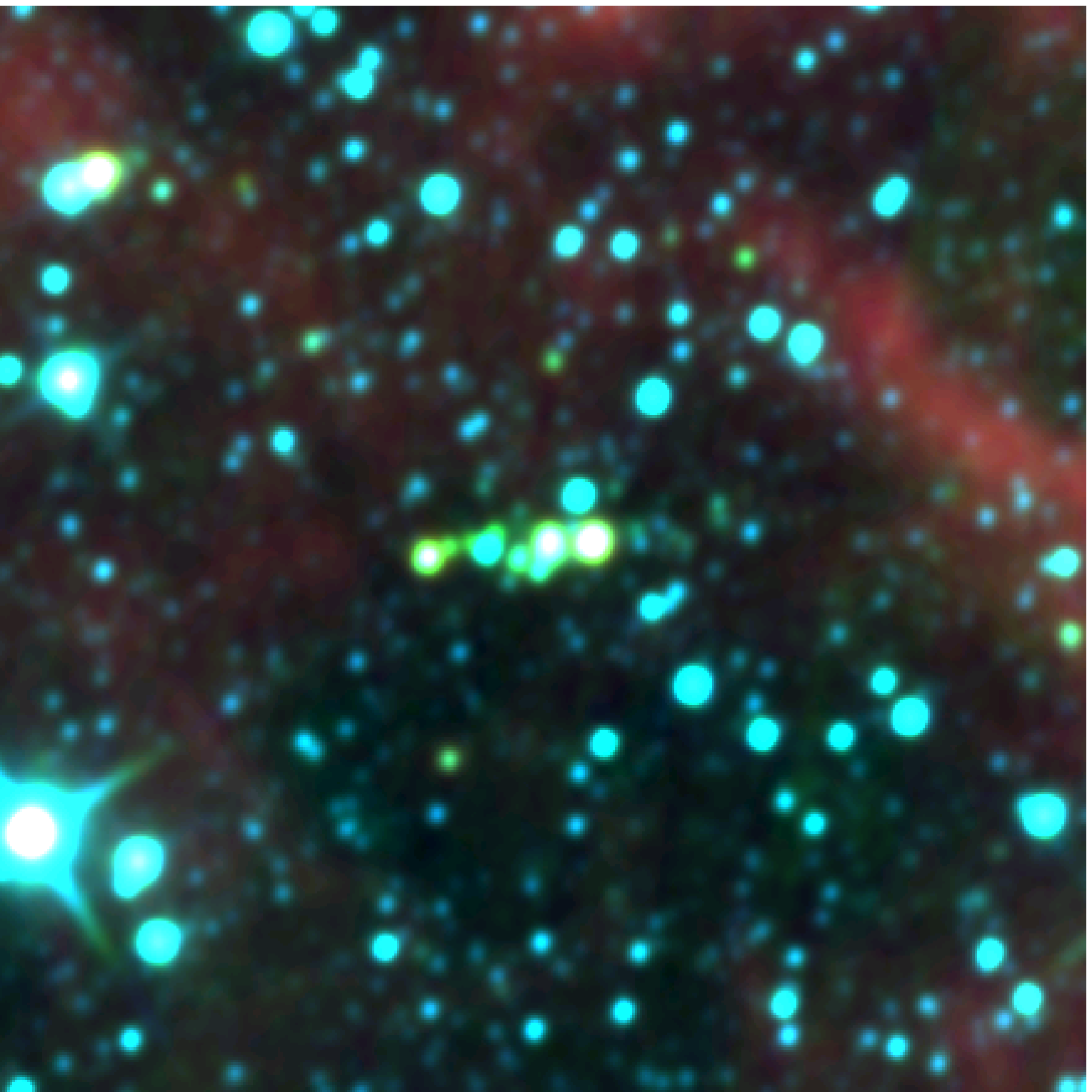}
\put(-140.0,155.0){\makebox(0.0,0.0)[5]{\fontsize{14}{14}\selectfont \color{red}C 444}}
\end{minipage}\hfill
%\hspace{0.01cm}
\begin{minipage}[b]{0.328\linewidth}
\includegraphics[width=\textwidth]{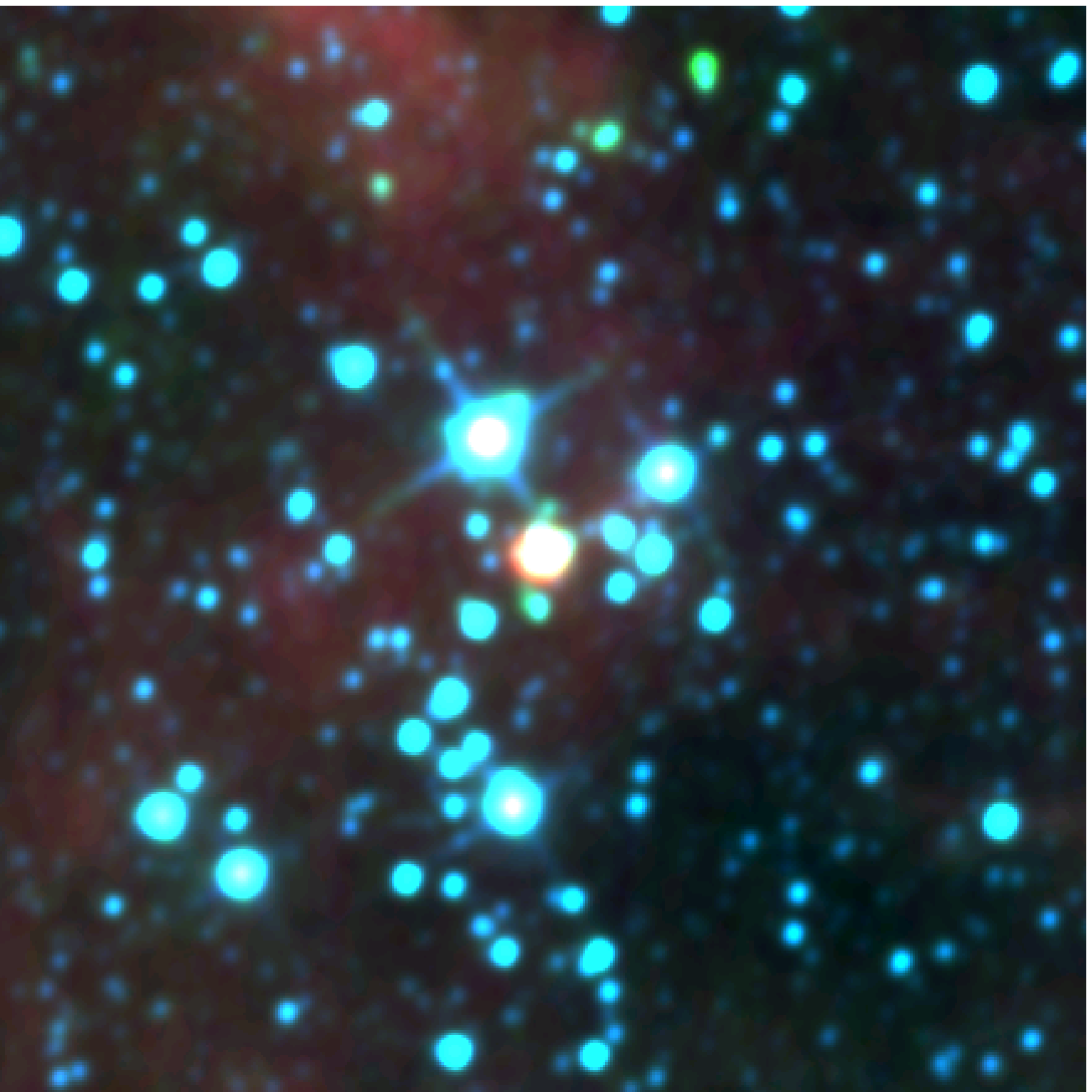}
\put(-140.0,155.0){\makebox(0.0,0.0)[5]{\fontsize{14}{14}\selectfont \color{red}C 443}}
\end{minipage}\hfill
\caption[]{WISE RGB images centred on FSR 665 ($10'\times10'$), Camargo 441 ($7'\times7'$), Camargo 442 ($12'\times12'$), FSR 666 ($10'\times10'$), Camargo 444 ($10'\times10'$), and Camargo 443 ($10'\times10'$).}
\label{f2}
\end{figure*}

The main goal of the present work is deriving accurate cluster parameters to use as tools to provide new constraints to better understand the Galactic structure.
We also present new ECs discovered by ourselves as a follow up of our recent catalogue \citep{Camargo15a, Camargo15b}. The searches for new clusters were made by eye on WISE images.

\begin{table*}
\centering
{\footnotesize
\caption{General data of the present star clusters.}
\label{tab1}
\renewcommand{\tabcolsep}{3.5mm}
\renewcommand{\arraystretch}{1,2}
\begin{tabular}{lrrrrr}
\hline
\hline
Target&$\alpha(2000)$&$\delta(2000)$&$\ell$&$b$&comments\\
&(h\,m\,s)&$(^{\circ}\,^{\prime}\,^{\prime\prime})$&$(^{\circ})$&$(^{\circ})$& \\
($1$)&($2$)&($3$)&($4$)&($5$)&($6$)\\
\hline
IRAS 0207+6047 Cl. &2:44:37&60:59:42&136.22&1.08& also known as BDB 116  and Majaess 30\\
SAI 23 &2:54:10&60:39:02&137.42&1.28&\citet{Glushkova10}\\
Camargo 440 &3:59:58&51:33:23&150.05&-1.11& present work\\
BDS 61 &4:03:17&51:19:35&150.59&-0.94&\citet{Bica03b}\\
Camargo 441 &4:03:18&51:29:37&150.48&-0.81& present work\\
Camargo 442 &4:04:14&51:22:56&150.66&-0.80& present work\\
FSR 665 &4:04:59&51:31:35&150.65&-0.62&\citet{Froebrich07}\\
FSR 666 &4:05:51&51:28:35&150.79&-0.56&\citet{Froebrich07}\\
Camargo 443 &4:07:08&51:11:14&151.13&-0.64&present work\\
Camargo 444 &4:07:49&51:15:23&151.16&-0.52&present work\\
Camargo 445 &4:08:12&50:31:29&151.70&-1.02& present work (in Majaess 45)\\
BDS 65 &4:11:10&51:09:58&151.61&-0.23&\citet{Bica03b}\\
Camargo 446 &6:10:27&16:43:14&193.43&-1.17&present work\\
Camargo 63 &6:12:01&13:39:36&196.30&-2.31&\citet{Camargo15a}\\
Majaess 78 &6:13:39&15:58:07&194.46&-0.87&\citet{Majaess13}\\
Dolidze 25 &6:45:02&00:13:21&212.00&-1.33& \citet{Moffat75}\\
Bochum 2 &6:48:50&00:22:44&212.29&-0.40&\citet{Turbide93}\\
NGC 2367 &7:20:07&-21:52:57&235.60&-3.83&\citet{Vogt72}\\
\hline
\end{tabular}
\begin{list}{Table Notes.}
\item Cols. $2-3$: Central coordinates. Cols. $4-5$: Corresponding Galactic coordinates. Col. $6$: comments.  
\end{list}
}
\end{table*}

The paper is organized as follows. In Sect. \ref{sec:2} we describe the methods and tools employed in the cluster analyses. In Sect. \ref{sec:3} we present the results of the cluster analysis, and derive parameters (\textit{age, reddening, distance, core and cluster radii}). In Sect. \ref{sec:4} we discuss the results. Finally, in Sect. \ref{sec:5} we provide the concluding remarks.

\section{Methods of analysis}
\label{sec:2}

Cluster fundamental parameters are derived  using 2MASS\footnote{The Two Micron All Sky Survey, available at \textit{www..ipac.caltech.edu/2mass/releases/allsky/}} photometry \citep{Skrutskie06} in the $J$, $H$ and $K_{s}$ bands, extracted in circular concentric regions centred on the coordinates given in Table~\ref{tab1} and fitted with PARSEC isochrones \citep{Bressan12}.
The fits are made by eye, taking the combined MS and PMS stellar distributions as constraints, and allowing for variations due to photometric uncertainties and differential reddening. 

\begin{figure}
\resizebox{\hsize}{!}{\includegraphics{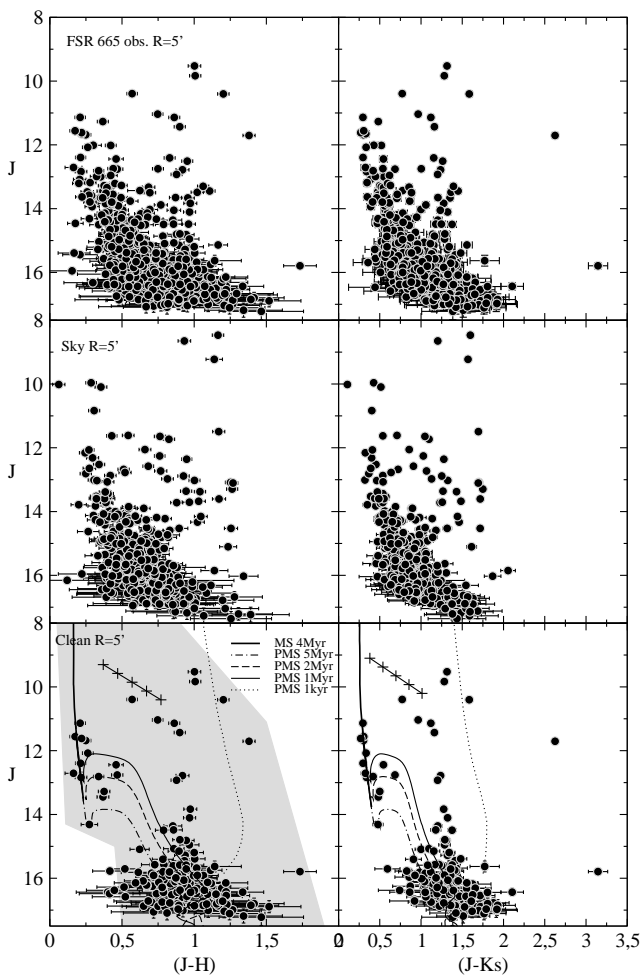}}
\caption[]{2MASS $J\times(J-H)$ and  $J\times(J-K_s)$ CMDs for FSR 665. \textit{Top panels}: observed CMDs. \textit{Middle}: equal area comparison field. \textit{Bottom}: field-star decontaminated CMDs. The decontaminated CMD of FSR 665 was fitted with PARSEC isochrones for both MS and PMS stars. The colour-magnitude filter used to isolate
cluster stars is shown as a shaded region. We also show the reddening vector for $A_V=0$ to 5.}
\label{f4}
\end{figure}

\begin{figure}
\resizebox{\hsize}{!}{\includegraphics{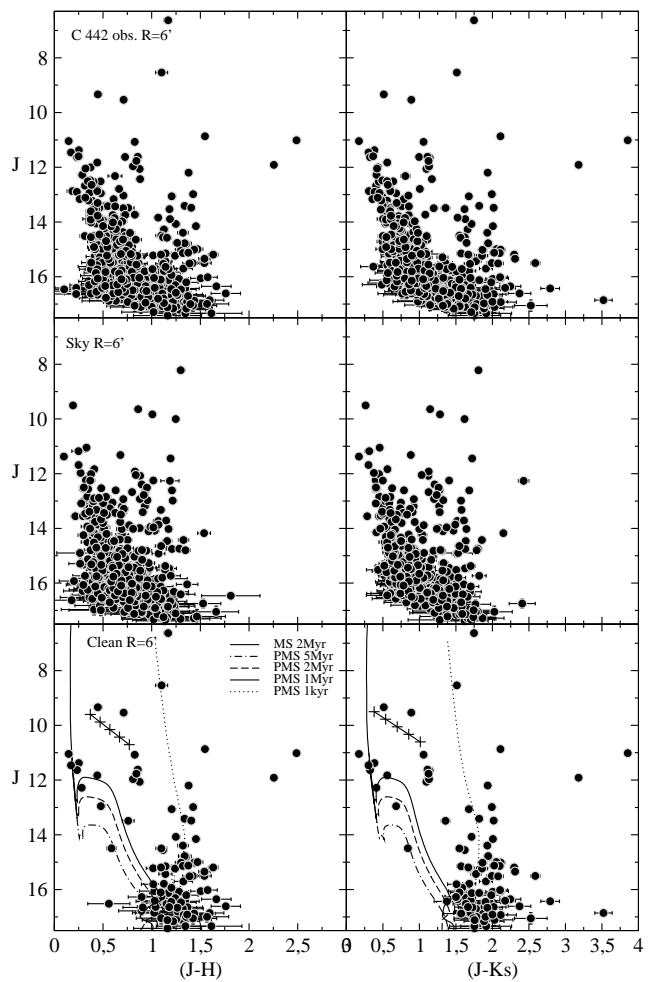}}
\caption[]{Same as Fig.~\ref{f4} for Camargo 442.}
\label{f5}
\end{figure}

\begin{figure}
\resizebox{\hsize}{!}{\includegraphics{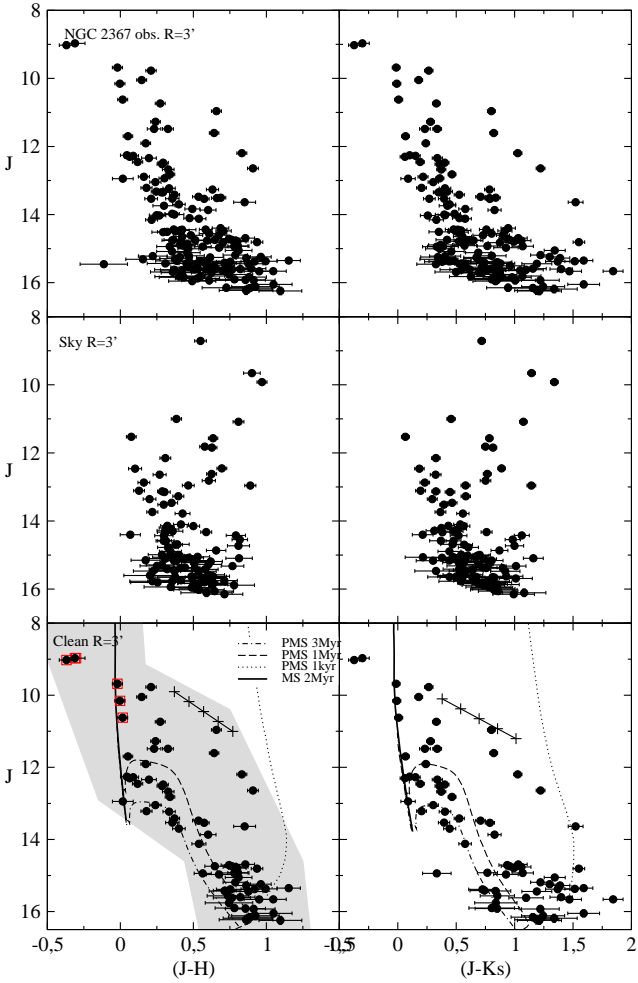}}
\caption[]{Same as Fig.~\ref{f4} for NGC 2367. The squares on the decontaminated CMDs indicate B stars.}
\label{f6}
\end{figure}

\begin{figure}
\resizebox{\hsize}{!}{\includegraphics{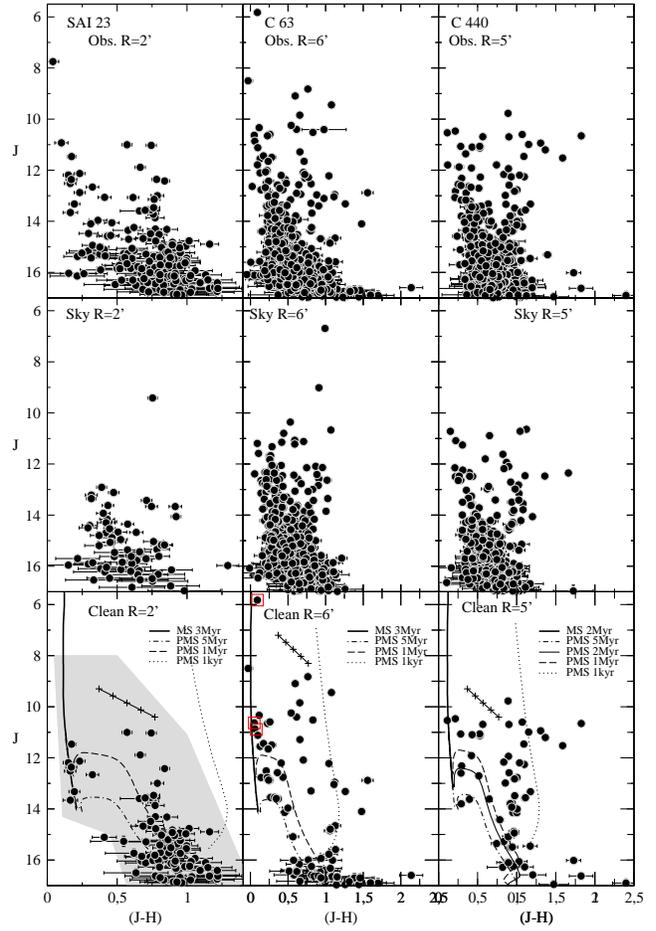}}
\caption[]{2MASS CMDs of the ECs SAI 23, Camargo 63, and Camargo 440. Top panels: observed CMDs $J\times(J-H)$. Middle panels: equal area comparison field. Bottom panels: field star decontaminated CMDs fitted with MS and PMS PARSEC isochrones. The colour-magnitude filters used to isolate cluster stars are shown as shaded regions (only for clusters with RDP following a King-like profile). The squares on the decontaminated CMDs indicate B stars. We also present the reddening vector for $A_V=0$ to 5.}
\label{fig:7}
\end{figure}

\begin{figure}
\resizebox{\hsize}{!}{\includegraphics{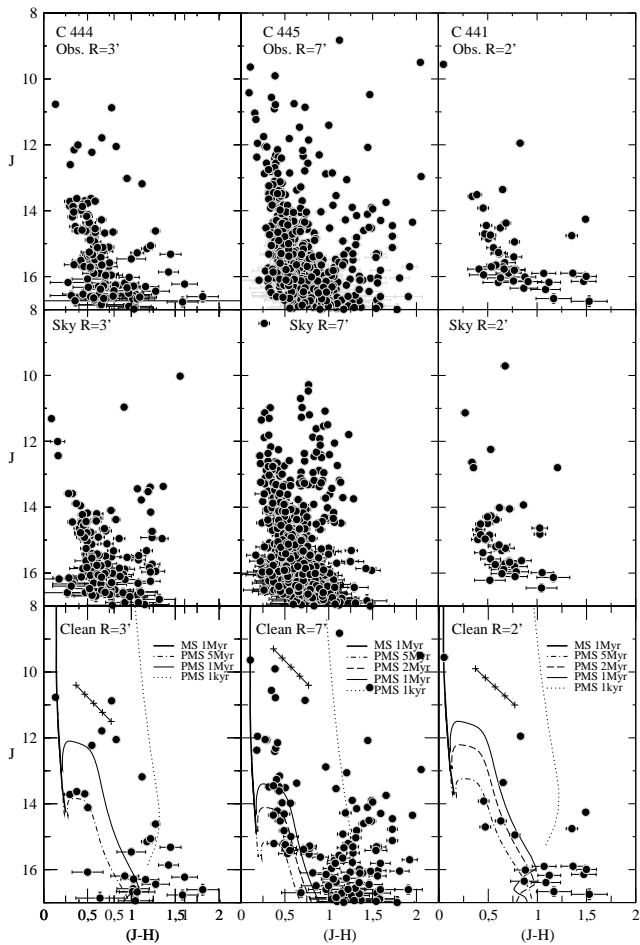}}
\caption[]{Same as Fig.~\ref{fig:7} for Camargo 444, Camargo 445, and Camargo 441.}
\label{fig:8}
\end{figure}

\begin{figure}
\resizebox{\hsize}{!}{\includegraphics{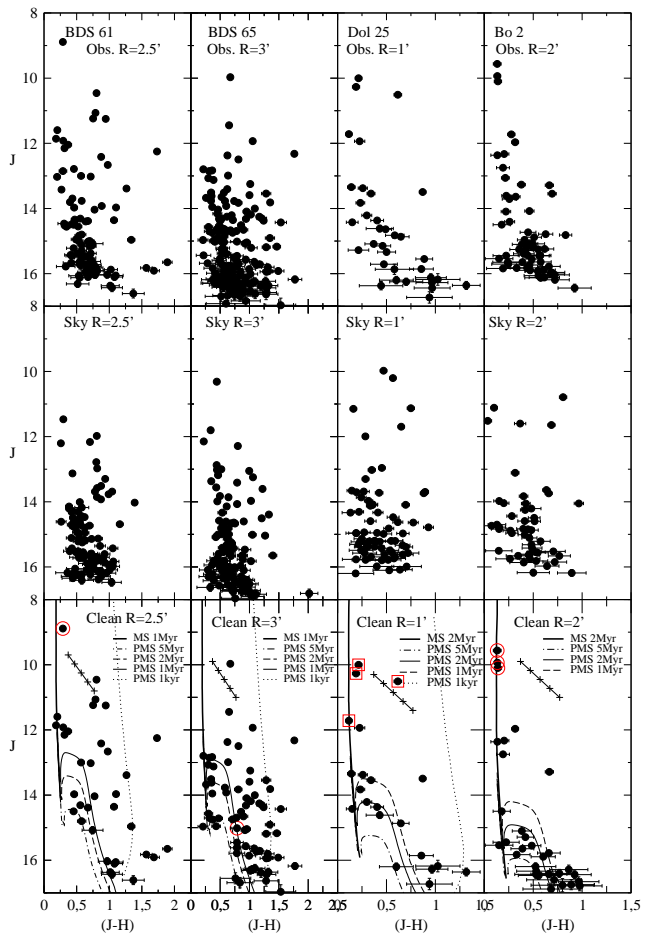}}
\caption[]{Same as Fig.~\ref{fig:7} for BDS 61, BDS 65, Dolidze 25, and Bochum 2. The circles on the decontaminated CMDs indicate O stars and the squares B stars.}
\label{fig:9}
\end{figure}

\begin{figure}
\resizebox{\hsize}{!}{\includegraphics{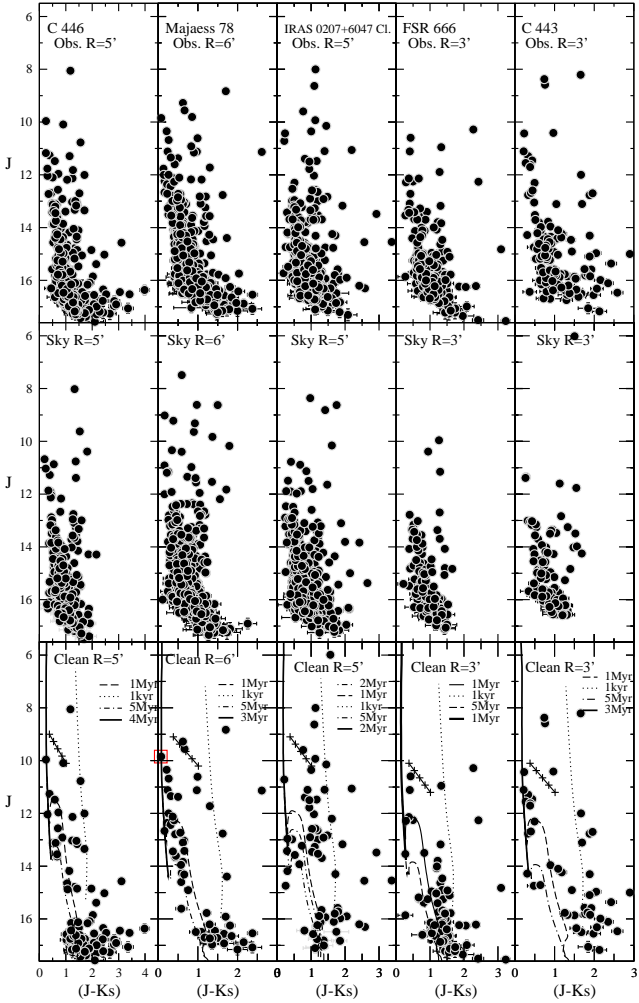}}
\caption[]{2MASS CMDs for the clusters Camargo 446, Majaess 78, IRAS $0207+6047$ Cl., FSR 666,  and Camargo 443. Top panels: observed CMDs $J\times(J-K_s)$. Middle panels: equal area comparison field. Bottom panels: field star decontaminated CMDs fitted with MS and PMS PARSEC isochrones. We also present the reddening vector for $A_V=0$ to 5.}
\label{fig:10}
\end{figure}

The process relies on application of shifts in magnitude and colour in the isochrone set (MS + PMS) from the zero distance modulus and reddening until a satisfactory solution is reached.
The best fits are superimposed on decontaminated CMDs (Figs.~\ref{f4} to \ref{fig:10}). 

To uncover the intrinsic CMD morphology, we apply a field-star decontamination procedure.  The algorithm works on a statistical basis by measuring the relative number densities of probable cluster and field stars in 3D CMD cells that have axes along the $J$, $(J - H)$, and $(J - K_s)$ magnitude and colours \citep{Bonatto07b, Bica08, Bonatto08, Bonatto10}. Then, the algorithm subtracts the expected number of field stars from each cell. 
It has been used in several works \citep[e.g.][and references therein]{Camargo09, Camargo10, Camargo11, Camargo12, Camargo13, Bica11, Bonatto09, Bonatto11a}.

The cluster structure is analysed by means of the stellar radial density profile (RDP) built with stars selected after applying the respective colour magnitude (CM) filter to the observed photometry.
The CM filter excludes stars with different colours of those within the probable cluster sequences and enhances the RDP contrast relative to the background \citep[e.g.][and references therein]{Bonatto07a}. Structural parameters are derived by fitting  King-like profiles to the clusters RDPs \citep{King62}. This procedure was applied in previous works \citep[e.g.,][and references therein]{Bica11, Bonatto10, Bonatto11b, Lima14}.

\begin{table*}
{\footnotesize
\begin{center}
\tiny{
\caption{Derived fundamental parameters for embedded clusters in the present study.}
\renewcommand{\tabcolsep}{0.5mm}
\renewcommand{\arraystretch}{1.3}
\begin{tabular}{lrrrrrrrrrrr}
\hline
\hline
Cluster&$E(J-H)$&Age&$d_{\odot}$&$R_{GC}$&$x_{GC}$&$y_{GC}$&$z_{GC}$&$R_{GC}$&$x_{GC}$&$y_{GC}$&$z_{GC}$\\
&(mag)&(Myr)&(kpc)&(kpc)&(kpc)&(kpc)&(pc)&(kpc)&(kpc)&(kpc)&(pc)\\
($1$)&($2$)&($3$)&($4$)&($5$)&($6$)&($7$)&($8$)&($9$)&($10$)&($11$)&($12$)\\
\hline
IRAS 0207+6047 Cl. &$0.28\pm0.02$&$2\pm1$&$2.8\pm0.4$&$9.4\pm0.3$&$-9.25\pm0.28$&$1.94\pm0.27$&$52.8\pm7.4$&$10.51\pm0.28$&$-10.33\pm0.28$&$1.94\pm0.27$&$52.89\pm7.42$\\
SAI 23 &$0.28\pm0.02$&$4\pm1$&$2.7\pm0.4$&$9.4\pm0.3$&$-9.19\pm0.28$&$1.81\pm0.25$&$59.8\pm8.4$&$10.43\pm0.28$&$-10.27\pm0.28$&$1.81\pm0.25$&$59.86\pm8.40$\\
Camargo 440 &$0.27\pm0.02$&$1\pm0.5$&$2.6\pm0.4$&$9.5\pm0.3$&$-9.46\pm0.31$&$1.29\pm0.18$&$-50.2\pm7.0$&$10.62\pm0.31$&$-10.54\pm0.31$&$1.29\pm0.18$&$-50.20\pm7.04$\\
BDS 61 &$0.23\pm0.03$&$2\pm1$&$2.7\pm0.3$&$9.7\pm0.3$&$-9.59\pm0.335$&$1.34\pm0.18$&$-44.7\pm6.2$&$10.76\pm0.33$&$-10.68\pm0.33$&$1.32\pm0.18$&$-21.27\pm2.95$\\
Camargo 441 &$0.21\pm0.03$&$1\pm0.5$&$2.5\pm0.5$&$9.5\pm0.4$&$-9.44\pm0.42$&$1.26\pm0.24$&$-36.0\pm6.8$&$10.59\pm0.41$&$-10.52\pm0.42$&$1.26\pm0.24$&$-36.01\pm6.76$\\
Camargo 442 &$0.32\pm0.02$&$1\pm1$&$2.7\pm0.4$&$9.6\pm0.3$&$-9.55\pm0.34$&$1.31\pm0.19$&$-37.3\pm5.5$&$10.71\pm0.34$&$-10.63\pm0.34$&$1.31\pm0.19$&$-37.27\pm5.47$\\

FSR 665 &$0.32\pm0.03$&$4\pm1.5$&$2.9\pm0.3$&$9.9\pm0.2$&$-9.77\pm0.2$&$1.43\pm0.14$&$-31.7\pm3.0$&$10.95\pm0.24$&$-10.85\pm0.24$&$1.43\pm0.14$&$-31.67\pm3.02$\\
FSR 666 &$0.27\pm0.03$&$1\pm1$&$2.8\pm0.7$&$9.8\pm0.6$&$-9.70\pm0.58$&$1.39\pm0.32$&$-27.9\pm6.5$&$10.87\pm0.57$&$-10.78\pm0.58$&$1.39\pm0.32$&$-27.77\pm6.47$\\
Camargo 443 &$0.30\pm0.02$&$3\pm1$&$3.1\pm0.7$&$10.1\pm0.6$&$-9.97\pm0.65$&$1.52\pm0.36$&$-35.1\pm8.3$&$11.16\pm0.64$&$-11.05\pm0.65$&$1.52\pm0.36$&$-35.10\pm8.26$\\
Camargo 444 &$0.29\pm0.03$&$2\pm1$&$3.0\pm0.7$&$10.0\pm0.6$&$-9.88\pm0.63$&$1.47\pm0.34$&$-27.6\pm6.5$&$10.93\pm0.35$&$-10.84\pm0.36$&$1.40\pm0.20$&$-26.34\pm3.69$\\
Camargo 445 &$0.26\pm0.02$&$1\pm0.5$&$6.3\pm0.9$&$13.1\pm0.7$&$-12.76\pm0.78$&$2.98\pm0.42$&$-112.1\pm15.7$&$14.16\pm0.76$&$-13.84\pm0.78$&$2.98\pm0.42$&$-112.05\pm15.71$\\
BDS 65 &$0.22\pm0.03$&$1\pm1$&$2.5\pm0.4$&$9.5\pm0.3$&$-9.43\pm0.33$&$1.20\pm0.18$&$-10.1\pm6.2$&$10.58\pm1.02$&$-10.51\pm1.02$&$1.20\pm0.55$&$-10.15\pm4.68$\\
Camargo 446 &$0.14\pm0.01$&$1\pm0.5$&$2.9\pm0.4$&$10.1\pm0.4$&$-10.05\pm0.39$&$-0.68\pm0.09$&$-59.4\pm8.2$&$11.15\pm0.39$&$-11.13\pm0.39$&$-0.68\pm0.09$&$-59.44\pm8.24$\\
Camargo 63  &$0.14\pm0.02$&$3\pm1$&$3.3\pm0.5$&$10.5\pm0.4$&$-10.43\pm0.45$&$-0.94\pm0.13$&$-131.2\pm18.4$&$11.54\pm0.45$&$-11.51\pm0.45$&$-0.94\pm0.13$&$-131.22\pm18.40$\\
Majaess 78 &$0.22\pm0.02$&$5\pm3$&$3.2\pm0.5$&$10.3\pm0.5$&$-10.29\pm0.50$&$-0.79\pm0.13$&$-47.5\pm7.7$&$11.39\pm0.50$&$-11.37\pm0.50$&$-0.79\pm0.13$&$-47.54\pm7.74$\\
Dolidze 25 &$0.25\pm0.01$&$2\pm1$&$5.5\pm0.8$&$12.3\pm0.7$&$-11.93\pm0.66$&$-2.94\pm0.66$&$-128.9\pm18.1$&$13.34\pm0.65$&$-13.01\pm0.66$&$-2.94\pm0.41$&$-128.85\pm18.07$\\
Bochum 2 &$0.30\pm0.01$&$5\pm1$&$7.9\pm1.1$&$14.5\pm0.9$&$-13.89\pm0.92$&$-4.22\pm0.58$&$-55.1\pm7.6$&$15.56\pm0.90$&$-14.97\pm0.92$&$-4.22\pm0.58$&$-55.11\pm7.64$\\
NGC 2367 &$0.12\pm0.01$&$2\pm1$&$3.3\pm0.7$&$9.5\pm0.4$&$-9.06\pm0.42$&$-2.69\pm0.62$&$-218.6\pm50.4$&$10.50\pm0.44$&$-10.14\pm0.42$&$-2.69\pm0.62$&$-218.60\pm50.35$\\

\hline
\end{tabular}
\begin{list}{Table Notes.}
\item Col. 2: E(B-V) in the cluster's central region. Col. 3: age, from 2MASS photometry. Col. 4: distance from the Sun. Col. 5: $R_{GC}$ calculated using $R_{\odot}=7.2$ kpc for the distance of the Sun to the Galactic centre \citep{Bica06}. Cols. 6 - 8: Galactocentric components using $R_{\odot}=7.2$ kpc. Col. 9: $R_{GC}$ calculated using $R_{\odot}=8.3$ kpc for the distance of the Sun to the Galactic centre. Cols. 10 - 12: Galactocentric components using $R_{\odot}=8.3$ kpc.
\end{list}
\label{tab2}}
\end{center}
}
\end{table*}

\begin{table*}
{\footnotesize
\begin{center}
\caption{Structural parameters for clusters in the current sample.}
\renewcommand{\tabcolsep}{3.5mm}
\renewcommand{\arraystretch}{1.3}
\begin{tabular}{lrrrrrrrr}
\hline
\hline
Cluster&$(1')$&$\sigma_{0K}$&$R_{core}$&$R_{RDP}$&$\sigma_{0K}$&$R_{core}$&$R_{RDP}$\\
&($pc$)&($*\,pc^{-2}$)&($pc$)&($pc$)&($*\,\arcmin^{-2}$)&($\arcmin$)&($\arcmin$)\\
($1$)&($2$)&($3$)&($4$)&($5$)&($6$)&($7$)&($8$)\\
\hline
SAI 23 &$0.77$&$25.23\pm3.63$&$0.85\pm0.10$&$10.01\pm2.31$&$14.96\pm2.15$&$1.11\pm0.13$&$13.0\pm3.0$\\
FSR 665 &$0.85$&$25.92\pm8.4$&$0.67\pm0.16$&$6.97\pm2.55$&$18.73\pm6.07$&$0.79\pm0.19$&$8.2\pm3.0$\\
Dol 25 &$1.61$&$21.42\pm5.98$&$0.31\pm0.06$&$8.05\pm3.22$&$55.52\pm15.50$&$0.19\pm0.04$&$5.0\pm2.0$\\
Bo 2 &$2.29$&$7.03\pm2.97$&$0.57\pm0.16$&$9.16\pm2.30$&$36.85\pm15.59$&$0.25\pm0.07$&$4.0\pm1.0$\\
NGC 2367 &$0.95$&$24.64\pm13.64$&$0.31\pm0.12$&$4.75\pm0.47$&$22.24\pm12.31$&$0.33\pm0.13$&$5.0\pm0.5$\\
\hline
\end{tabular}
\label{tab3}
\end{center}
}
\end{table*}

\begin{figure}
\resizebox{\hsize}{!}{\includegraphics{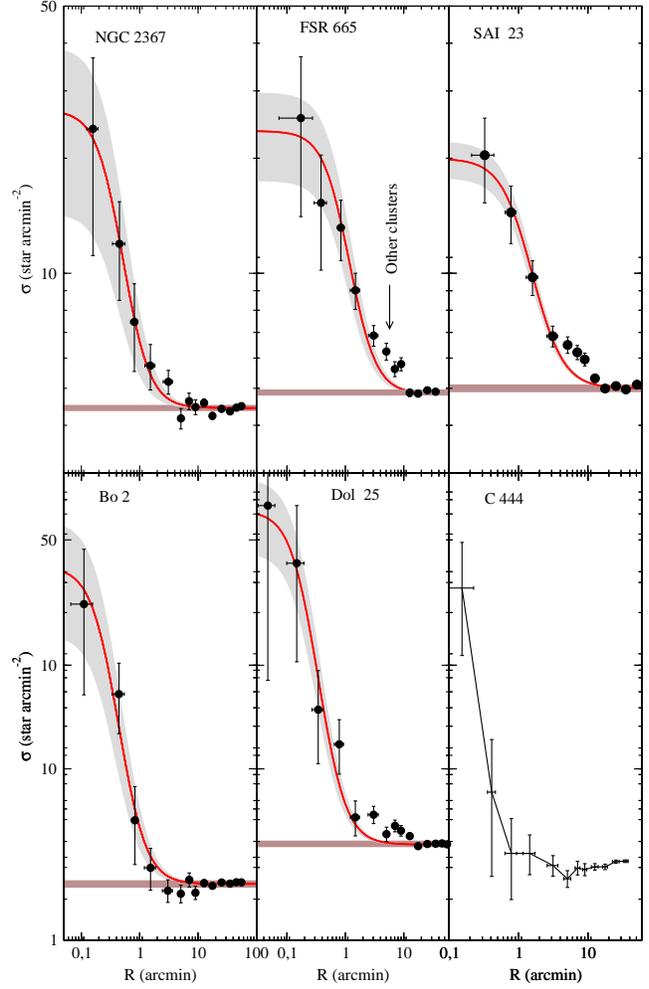}}
\caption[]{Radial density profiles for the ECs NGC 2367, FSR 665, SAI 23, Bochum 2, Dolidze 25, and Camargo 444. Brown horizontal shaded region: stellar background level measured in the comparison field. Gray regions: $1\sigma$ King fit uncertainty.}
\label{rdp}
\end{figure}

\begin{figure}
\resizebox{\hsize}{!}{\includegraphics{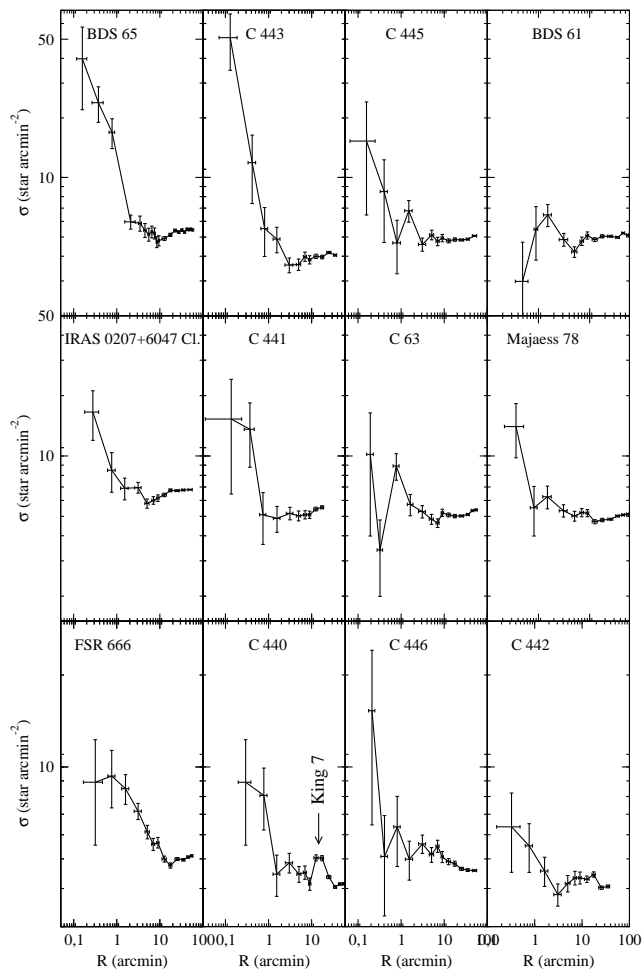}}
\caption[]{Radial density profiles for the confirmed clusters BDS 65, Camargo 443, Camargo 445, BDS 61, IRAS 0207+6047 Cl., Camargo 441, Camargo 63, Majaess 78, FSR 666, Camargo 440, Camargo 446, and Camargo 442.}
\label{rdp2}
\end{figure}

\section{Results}
\label{sec:3}

NASA's WISE telescope observed the whole sky in the bands W1 ($3.4{\mu}m$), W2($4.6{\mu}m$), W3($12{\mu}m$) and W4 ($22{\mu}m$). W1 and W2 are particularly sensitive to the EC stellar content, while W3 and W4 show mostly dust emission.

Figs.~\ref{nuvem} and  \ref{f2} show WISE composite images for BDS 61, BDS 65, C 441, C 442, C 443, C 444, FSR 665, and FSR 666. Figs.~\ref{f4} to \ref{fig:10} show CMDs for all objects in Table~\ref{tab1}. RDPs are presented in Figs.~\ref{rdp} and \ref{rdp2}. Fundamental parameters are shown in Table~\ref{tab2} and the structure for clusters that can be described by a King law are shown in Table~\ref{tab3}.

\subsection{Previously known star clusters}
\label{sec:3.1}

IRAS 0207+6047 Cluster (BDB 116): we derived an age of $2\pm1$ Myr for a distance of $d_{\odot}=2.8\pm0.4$ kpc (Fig.~\ref{fig:10}). The structure of this object points to a cluster, but cannot be fitted by a King-like function (Fig.~\ref{rdp2}).

SAI 23: fitting isochrones to cluster decontaminated CMD (Fig.~\ref{fig:7}), we derive an age of $\sim4$ Myr for a distance of $d_{\odot}=2.8\pm0.4$ kpc. From the RDP (Fig.~\ref{rdp}) we found $R_{c}=0.85\pm0.1$ pc, $\sigma_{0K}=25.2\pm3.6\,stars\,pc^{-2}$, and a cluster RDP radius of  $10.01\pm2.3$ pc.

BDS 61: is an EC located in the Perseus arm presenting an O type star in the central region ($R<1'$), as shown by the decontaminated CMD (Fig.~\ref{fig:9}). We derive an age of $\sim1$ Myr for a $d_{\odot}=2.7\pm0.3$ kpc. Despite a dip in the inner RDP region, the structure of BDS 61 is consistent with an EC in the early evolutionary phase (Fig.~\ref{rdp2}).  

FSR 665: located close to the Perseus arm, this object is a prominent EC with a relatively well-developed MS and a populous PMS (Fig.~\ref{f4}). The best fit of PARSEC isochrones (MS + PMS) suggest an age of $\sim4$ Myr and a $d_{\odot}=2.9\pm0.3$ kpc. The presence of other ECs creates a bump in the FSR 665 RDP (Fig.~\ref{rdp}).

FSR 666: \textit{WISE} images (Figs.~\ref{nuvem} and \ref{f2}) of FSR 666 point to ongoing star formation. The decontaminated CMD and the respective MS and PMS isochrone fits provide an age of $\sim1$ Myr for a $d_{\odot}=2.8\pm0.7$ kpc, which put this EC in the Perseus arm. The RDP of FSR 666 is irregular and cannot be fitted by a King-like profile (Fig.~\ref{rdp2}), but suggests a cluster. Note
that young clusters lack the span of time to be an isothermal sphere.

BDS 65: located in the Perseus arm at a $d_{\odot}=2.5\pm0.4$ kpc. The object is a prominent EC, as indicated by CMDs (Fig.~\ref{fig:9}) and RDP (Fig.~\ref{rdp2}). The decontaminated CMD of BDS 65 shows a very reddened O star within the central region ($R<1'$) and suggests an age of $\sim1$ Myr. The RDP presents a high contrast with respect to the background, but King profile cannot be fitted.

Camargo 63: recently discovered by \citet{Camargo15a} this EC is confirmed as a cluster in the present work. The decontaminated CMD of C 63 fitted by PARSEC isochrones (Fig.~\ref{fig:7}) suggests an age of $\sim3$ Myr and present  3 B stars in the central region. We found a distance from the Sun of $3.3\pm0.5$ kpc, close to the Perseus arm. The RDP of C 63 is irregular (Fig.~\ref{rdp2}), certainly owing to absorption effects in the 2MASS bands.

Majaess 78: our analyses (Fig.~\ref{fig:10}) suggest that this EC lies in the Perseus arm at a distance of $3.2\pm0.5$ kpc and presents an age of $\sim3$ Myr. Fig.~\ref{rdp2} shows the irregular RDP for Majaess 78.

Dol 25: \citet{Moffat75} derived a distance from the Sun of $5.25$ kpc and \citet{Lennon90} a $d_{\odot}=5.5\pm0.5$ kpc. We argue that Dol 25 is related to Sh2-284 and lies at $d_{\odot}=5.5\pm0.8$ kpc. We obtained an age of $\sim2$ Myr for Dol 25. Five B-type stars are found in the central region of the cluster. The inner region of the cluster ($R<1'$) contains four of the massive B stars shown in the decontaminated CMD (Fig.~\ref{fig:9}). The RDP fitted by a King-like profile (Fig.~\ref{rdp}) gives $R_{c}=0.31\pm0.06$ pc, $\sigma_{0K}=21.42\pm5.98\,stars\,pc^{-2}$, and $R_{RDP}=8.05\pm3.22$ pc. The RDP irregularities at large radii are probably caused by stellar overdensities in the Dol 25 neighbourhood, probably related to sequential star formation.

Bochum 2: \citet{Stephenson71} found a distance from the Sun of $5.5$ kpc for Bo 2. subsequently, \citet{Moffat79} estimated a distance of $4.8$ kpc. \citet{Munari95} estimated an age of 7 Myr for a distance from Sun of $\sim6$ kpc. Our analyses of the decontaminated CMD of Bo 2 (Fig.~\ref{fig:9}) suggest an age of $\sim5$ Myr and a $d_{\odot}=7.9\pm1.1$ kpc. This EC lies in the Outer arm.  The O stars shown in the decontaminated CMD are members of binary systems and are located in the inner cluster radius ($R<1'$). The RDP of Bo 2 provides $R_{c}=0.57\pm0.16$ pc, $\sigma_{0K}=7\pm3\,stars\,pc^{-2}$ for a cluster radius of $9.2\pm2.3$ pc.

NGC 2367: \citet{Carraro05} derived an age of 5 Myr and a distance from the Sun of 1.4 kpc for NGC 2367 while \citet{Santos12} derived an age of 3 Myr and $d_{\odot}\sim2.2$ kpc. We estimated an age of $2\pm1$ Myr and $d_{\odot}=3.4\pm0.7$ kpc for NGC 2367. It presents five B stars in the central region, two of them with J excess (Fig.~\ref{f6}). The cluster shows a smooth RDP (Fig.~\ref{rdp}) with  $\sigma_{0K}=24.64\pm13.64\,stars\,pc^{-2}$ and $R_{RDP}=4.75\pm0.47$ pc. Both, CMDs and RDP suggest that this EC will evolve to become a classical OC.

\subsection{Present work discoveries}
\label{sec:3.2}

We discovered in this work 7 new ECs. The newly-found star clusters complement our previous catalogue \citep{Camargo15a, Camargo15b}.

Camargo 440: this new EC is located in the Perseus arm at a $d_{\odot}=2.6\pm0.4$ kpc. Based on the cluster CMD (Fig.~\ref{fig:7}), we estimated an age of $\sim1$ Myr. The overdensity in the outermost RDP of C 440 is due to King 7 (Fig.~\ref{rdp2}).

Camargo 441: this EC present an age of 1 Myr and is located at a $d_{\odot}=2.5\pm0.5$ kpc in the Perseus arm. Both the decontaminated CMD (Fig.~\ref{fig:9}) and RDP (Fig.~\ref{rdp2}) suggest a small poor cluster.

Camargo 442: this EC lies in the Perseus arm at a $d_{\odot}=2.7\pm0.4$ kpc. The best isochrone fit (Fig.~\ref{f5}) provides an age of $\sim1$ Myr. The irregular RDP (Fig.~\ref{rdp2}) does not follow a cluster-like profile.

Camargo 443: this  previously unknown EC lies in the Perseus arm at a distance of  $d_{\odot}=3.1\pm0.7$ kpc. C 443 is $\sim3$ Myr old and  presents a well-defined MS and PMS with very reddened stars (Fig.~\ref{fig:10}).  Its RDP suggests a cluster, but does not follow a King-like profile (Fig.~\ref{rdp2}), which is expected for young clusters \citep{Camargo15a}.

Camargo 444: lies at a distance of $d_{\odot}=3.0\pm0.7$ kpc in the Perseus arm. The analysis of the images (Figs.~\ref{nuvem} and \ref{f2}) and decontaminated CMD (Fig.~\ref{fig:8}) suggests ages of $\sim1-2$ Myr. The RDP (Fig.~\ref{rdp}) points to cluster, but cannot be fitted by King-like profile.

Camargo 445: we derive via the decontaminated CMD (Fig.~\ref{fig:8}) analysis an age of $\sim1$ Myr and a $d_{\odot}=6.3\pm0.9$ kpc, which sets this EC in the Outer arm. The RDP of C 445 is irregular and cannot be fitted by a King-like profile (Fig.~\ref{rdp2}).

Camargo 446: this object is an EC with an age of $\sim1$ Myr and is located in the Perseus arm at a distance of $d_{\odot}=2.9\pm0.4$ kpc (Fig.~\ref{fig:10}). The RDP does not follow a King law (Fig.~\ref{rdp2}).

\section{Discussion}
\label{sec:4}

The present analysis places the ECs BDS 61, C 441, C 442, C 443, C 444, FSR 665, and FSR 666 in the same region in the Perseus arm. 
Fig.~\ref{nuvem} presents this star cluster aggregate (hereafter Perseus 1 cluster aggregate) formed by ECs with similar ages. Such groups were predicted or observed by \citet{Efremov95, Guillout98, Fellhauer05, Fuente08, Fuente09a, Fuente10, Feigelson11, Camargo11, Camargo12, Camargo13, Camargo15a}.  Several large dust emission bubbles in WISE images connect these clusters, suggesting second generation effects \citep{Camargo15a}. This leaves open the possibility that an entire giant molecular cloud (GMC) may fragment almost simultaneously forming a large EC aggregate. The spiral arms may play an important role in the erosion of a GMC triggering massive star formation in the whole cloud. \citet{Camargo12} pointed out the star formation within cluster groups. They suggest that an irregular GMC (or complex) may form massive stars simultaneously and their winds and/or supernova explosions may produce a second generation of massive stars propagating the star formation and producing star clusters with similar age. 
The ECs formation is quite rapid and probably range from $0.5-5$ Myr \citep{Lada03, Allen07, Tamburro09, Santos12, Camargo13}. According to \citet{Camargo13},  the deep embedded phase for most Galactic ECs ends before 3Myr \citep{Ballesteros07, Fuente09b}.
 \citet{Bally08} argue that OB associations consist of sub-groups, clusters, and sub-clusters with no preferred scale \citep{Jose13}. The  GMC W51 with an EC aggregate distributed inside $\approx100$pc may be an example of this scenario \citep{Kumar04}. Several H{\sc ii} regions in \citet{Anderson14} are within complexes. Several EC candidates from \citet{Solin14}  have neighbours. The present Perseus 1 aggregate of ECs is similar to the Auriga 1 aggregate, recently found by \citet{Camargo12}.

The large-scale cluster formation within the structure shown in Fig.~\ref{nuvem} may be a consequence of the Perseus arm gas compression, but the small-scale cluster distribution suggest sequential formation. 

Dol 25 is located in the centre of a large bubble and apparently is triggering a sequential star formation event, forming a second EC generation \citep{Lee12, Camargo15a}. In this sense, EC aggregates appear to be formed in sequential events.  

BDS 61, BDS 65, Dol 25, and Bochum 2 present massive stars located within their central region ($R<1'$).

\subsection{Galactic distribution}
\label{sec:4.2}

\begin{figure*}
\begin{center}
   \includegraphics[scale=0.70,angle=0]{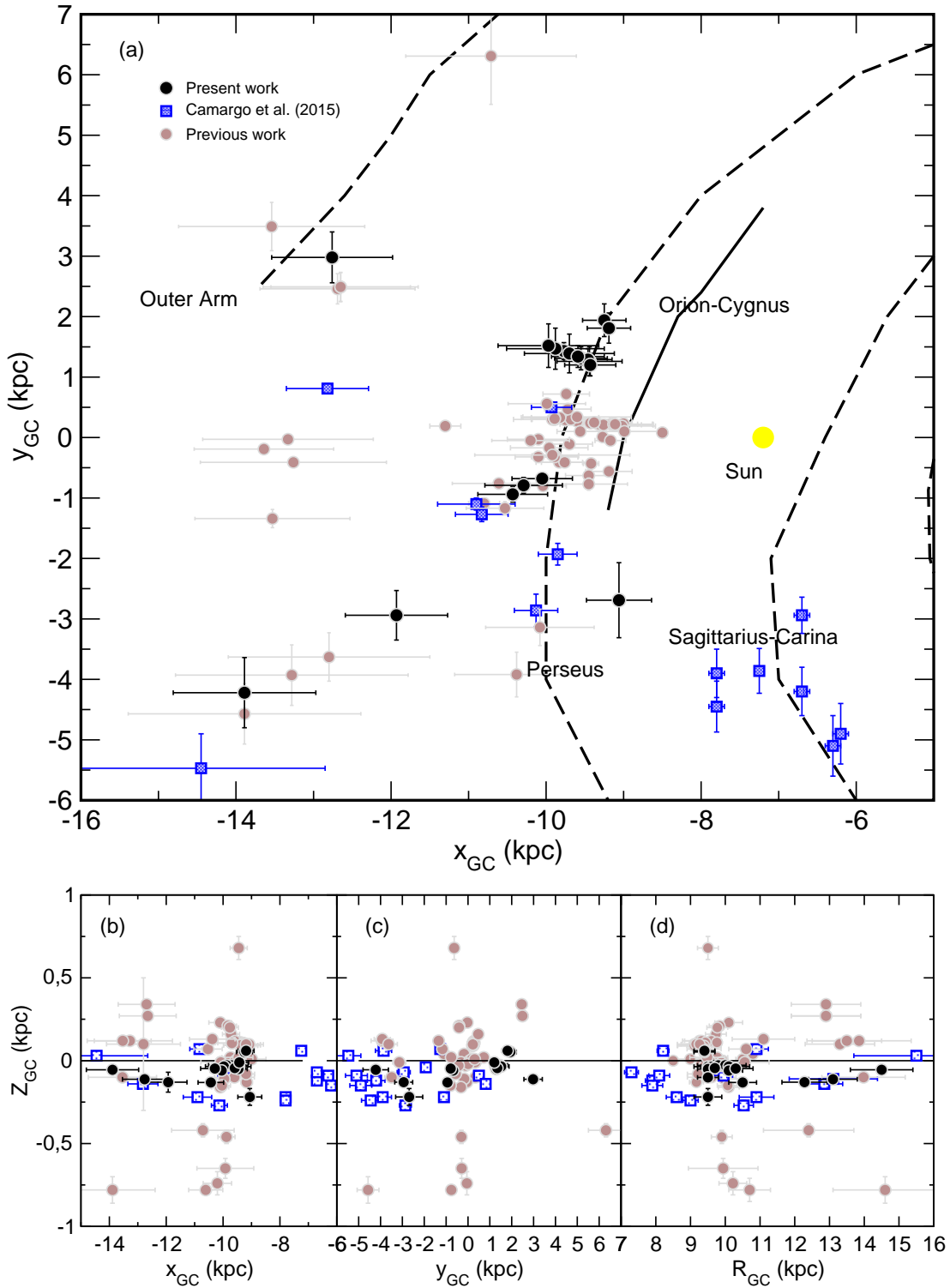}
   \caption[]{Spatial distribution of the ECs in schematic projection of the Galaxy as seen from the north pole, with 7.2 kpc as the Sun’s distance to the Galactic centre. Black circles are ECs in the present work, squares are ECs from \citet{Camargo15a}, and brown circles are ECs in our previous works.}
\label{fig:14}
\end{center}
\end{figure*}

Fig.~\ref{fig:14} updates the previous version in Fig. 14 in \citet{Camargo13}. Black circles are clusters in the present work and brown circles ECs from our previous studies \citep{Camargo10, Camargo11, Camargo12, Camargo13, Camargo15a}.

Following \citet{Camargo13}, most ECs are found in the thin disc within 250 pc from the Galactic mean plane \citep{Vallenari00, Siebert03} as shown in panels (b), (c), and (d).
The $Z_{GC}$ distribution of the present cluster sample (Table~\ref{tab2}) is consistent with \citet{Buckner14} result for the Galactic cluster scale height. However, some ECs are found at large distances from the Galactic plane (Fig.~\ref{fig:14}), mainly in the spiral arms \citep{Camargo13}. 

The Sagittarius-Carina spiral arm is well traced by the subsample of clusters with derived parameters in our recent discovered ECs \citep{Camargo15a}. This spiral arm is presently probed near the borderline of the third and fourth quadrants at a distance from the Galactic centre of  $d_1\sim6.4$ kpc  or $d_2\sim7.2$ kpc.

Most ECs in the present sample are distributed in the second and third quadrants along the Perseus arm. In this region the Perseus arm is located at  Galactocentric distances in the range of 9 kpc in the second quadrant to 10.5 kpc in the third quadrant for a distance of the Sun to the Galactic centre of 7.2 kpc or in the range of 9.8 to 11.3 kpc for $R_{\odot}=8.0$ kpc.

The presence of molecular clouds beyond the Perseus arm is known since a few decades \citep{Digel94, Heyer98}. \citet{Kaltcheva00} proposed the existence of a spiral arm in the Outer Galaxy, the Outer arm. Since then, several works have been developed, but the Outer arm is still not fully traced  \citep[][and references therein]{Russeil03, Pandey06, Honma07, Russeil07, Hachisuka09}. In \citet{Camargo13}, based on the distribution of ECs, we confirmed that the Outer arm extends along the second and third Galactic quadrants with Galactocentric distances in the range of $12.5-14.5$ kpc for $R_{\odot}=7.2$ kpc or $13.5-15.5$ kpc for $R_{\odot}=8.0$ kpc. The distance from the Sun and Galactocentric radii derived for the Outer arm in the present work agrees with \citet{Hachisuka09} results. There is a large discrepancy between the stellar Outer arm and the gaseous Outer arm with distance larger than 20 kpc, but it appears to be a common feature for large spiral galaxies \citep{McClure04}.

\section{Concluding remarks}
\label{sec:5}

In this work, we investigate the properties of 18 ECs.  Besides 11 previously known clusters we discovered 7 ECs, some of them forming a prominent EC aggregate located in the Perseus arm. The present results indicate that in the Galaxy ECs are predominantly located in the spiral arms. Thus, new searches for ECs are important, since they may contribute to the further understanding of the Galaxy structure.

The ECs in the present sample are distributed along three arms, Sagitarius-Carina, Perseus, and Outer arm.

The Sagittarius-Carina spiral arm in the region traced by our EC sample is at a Galactocentric distance of $\sim6.4$ kpc for $R_{\odot}=7.2$ kpc or $\sim7.2$ kpc for $R_{\odot}=8.0$ kpc.

The Perseus arm along the second and third quadrants presents a Galactocentric distance in the range 9 to 10.5 kpc for a distance of the Sun to the Galactic centre of 7.2 kpc or in the range of 9.8 to 11.3 kpc for a $R_{\odot}=8.0$ kpc. 

The parameters derived for Bochum 2 and C 445 reinforce our previous results for the Outer arm. This feature extends along the second and third Galactic quadrants with a distance from the Galactic centre in the range of $12.5-14.5$ kpc for $R_{\odot}=7.2$ kpc or $13.5-15.5$ kpc for $R_{\odot}=8.0$ kpc.
 
We find that in the Galaxy most ECs are distributed within the thin disc ($\sim250$ pc) and along spiral arms. However, there occur ECs with $z_{GC}>500$ pc.

Most massive stars identified in the present EC sample are located within the central region of the cluster ($R<1'$).

\vspace{0.8cm}

\textit{Acknowledgments}: We thank an anonymous referee for important comments and suggestions.  This publication makes use of data products from the Two Micron All Sky Survey, which is a joint project of the University of Massachusetts and the Infrared Processing and Analysis Centre/California Institute of Technology, funded by the National Aeronautics and Space Administration and the National Science Foundation. C.B. and E.B. acknowledge support from CNPq (Brazil).

\label{lastpage}
\end{document}